\documentclass[epsf,prb,twocolumn,nofootinbib]{revtex4-1}

\usepackage[pdftex]{graphicx}
\usepackage{dcolumn}
\usepackage{bm}
\usepackage{epsfig}
\usepackage{latexsym}
\usepackage{amsmath}
\usepackage{amsfonts}
\usepackage{amssymb}
\usepackage{color}
\usepackage{array}
\usepackage{framed}

\begin{document}

\title{The Fracton Gauge Principle}
\author{Michael Pretko \\
\emph{Department of Physics and Center for Theory of Quantum Matter \\ University of Colorado, Boulder, CO 80309}}
\date{\today}

\begin{abstract} 
A powerful mechanism for constructing gauge theories is to start from a theory with a global symmetry, then apply the ``gauge principle," which demands that this symmetry hold locally.  For example, the global phase rotation of a system of conserved charges can be promoted to a local phase rotation by coupling to an ordinary $U(1)$ vector gauge field.  More recently, a class of particles has been studied featuring not only charge conservation, but also conservation of higher moments, such as dipole moment, which leads to severe restrictions on the mobility of charges.  These particles, called fractons, are known to be intimately connected to symmetric tensor gauge fields.  In this work, we show how to derive such tensor gauge theories by applying the gauge principle to a theory of ungauged fractons.  We begin by formulating a field theory for ungauged fractons exhibiting global conservation of charge and dipole moment.  We show that such fracton field theories have a characteristic non-Gaussian form, reflecting the fact that fractons intrinsically interact with each other even in the absence of a mediating gauge field.  We then promote the global higher moment conservation laws to local ones, which requires the introduction of a symmetric tensor gauge field.  Finally, we extend these arguments to other types of subdimensional particles besides fractons.  This work offers a possible route to the formulation of non-abelian fracton theories.
\end{abstract}
\maketitle

\normalsize

\emph{Introduction}.  The concept of a gauge theory is one of the most important elements of the toolbox of modern theoretical physics, describing phenomena ranging from the fundamental particles of the Standard Model to topological phases of matter in solid state systems.  Gauge theories are different from more conventional field theories, in that they have an enormously large degree of symmetry, partially due to a significant redundancy in the description of physical states.  Specifically, gauge theories have \emph{local} symmetries, involving independent symmetry transformations at each point in spacetime, as opposed to the global symmetry transformations encountered in ungauged field theories.  A particularly elegant and powerful approach to constructing gauge theories is to start from an ungauged theory, with some global symmetry, then make the demand that this symmetry should continue to hold locally.  Philosophically, the notion that a global symmetry should continue to hold at the local level is known as the ``gauge principle."\cite{spell}

As a concrete example, consider a system of conserved particles, described by a field $\Phi$.  Particle number conservation will be encoded in the theory in the form of invariance of the action under a global phase rotation:
\begin{equation}
\Phi\rightarrow e^{i\alpha}\Phi
\end{equation}
for constant $\alpha$.  In order to construct a gauge theory, we now demand that the theory should be invariant under independent phase rotations at each point in spacetime, $i.e.$ under $\Phi\rightarrow e^{i\alpha(x,t)}\Phi$ for a function $\alpha(x,t)$ with arbitrary spacetime dependence.  Adding such spacetime dependence creates problems for derivative operators in the theory, which no longer transform covariantly:
\begin{equation}
\partial_\mu\Phi\rightarrow e^{i\alpha}(\partial_\mu + i\partial_\mu\alpha)\Phi
\end{equation}
In order to restore gauge-covariance and thereby obtain a gauge-invariant action, we must introduce a gauge field to absorb the non-covariant piece of this transformation.  We do this by defining a gauge-covariant derivative as:
\begin{equation}
D_\mu\Phi = (\partial_\mu - i A_\mu)\Phi\rightarrow e^{i\alpha}D_\mu\Phi
\end{equation}
where the gauge field $A_\mu$ must transform as $A_\mu\rightarrow A_\mu + \partial_\mu\alpha$.  We can then easily write down a gauge-invariant Lagrangian for the theory in terms of the gauge covariant derivative and field invariants of the gauge field itself:
\begin{equation}
\mathcal{L} = \mathcal{L}_m[\Phi,D\Phi,D^2\Phi,...] + F_{\mu\nu}F^{\mu\nu}
\end{equation}
where the first term describes matter and its interaction with the gauge field, while $F_{\mu\nu} = \partial_\mu A_\nu - \partial_\nu A_\mu$ is the standard field strength tensor of a $U(1)$ gauge field, describing the dynamics of a photon mode.  In this way, promoting the phase rotation of a set of conserved particles from a global to a local symmetry has led to the familiar Maxwell gauge theory.

Taking other types of global symmetries and ``gauging" them will result in other types of gauge theories.  For example, gauging a global symmetry described by a nonabelian Lie group will lead to a Yang-Mills theory.\cite{spell}  As another example, gauging a symmetry protected topological (SPT) phase protected by group $G$ will result in a topologically ordered phase with gauge group $G$.\cite{levin}  These types of gauging procedures are now well-established.  In recent years, however, a new set of unfamiliar symmetries and conservation laws has come to light, pertaining to higher charge moments.  These new conservation laws are manifested in systems of particles known as fractons, which have severely restricted mobility tracing directly to conservation of quantities such as dipole moment.\cite{sub}  (We refer the reader to a recent review\cite{review} for a broad perspective on fractons, and to selected literature\cite{sub,chamon,haah,fracton1,fracton2,genem,higgs1,higgs2,field,generic,
mach,holo,glassy,spread,deconfined,matter,sagarlayer,hanlayer,entanglement,
albert,elasticity,pai,fractalsym} for further details.)  While global charge conservation is associated with a global phase rotation, $\Phi\rightarrow e^{i\alpha}\Phi$, global dipole conservation is associated with a linearly varying phase rotation, $\Phi\rightarrow e^{i\vec{\lambda}\cdot \vec{x}}\Phi$.  This symmetry transformation is neither a gauge symmetry nor one of the recently discussed subsystem symmetries.\cite{fracton2,williamson,subsystem}  (Note that a subsystem symmetry is equivalent to an infinite number of higher moment symmetries, which provide the more general framework.)  It is not immediately clear what sort of theory will result from ``gauging" such a set of symmetries and conservation laws.  A significant clue is provided by fracton phases in certain spin, Majorana, and quantum rotor systems\cite{haah,fracton1,fracton2,sub,genem,higgs1,higgs2}, which are often described in the language of tensor gauge theories, as opposed to the more familiar vector gauge fields obtained from gauging conventional symmetries.  In this work, we will demonstrate explicitly how tensor gauge fields arise via application of the gauge principle to a system of fractons, promoting global higher moment conservation laws and symmetries to local ones.  This work serves as a useful complement to the existing literature on gauging subsystem symmetries to obtain discrete fracton models.\cite{fracton2,williamson,field,higgs1,higgs2,ungauge,fractalsym,foliated}

In order to study gauging a system of fractons, it is first necessary to have a field theory for ungauged $U(1)$ fractons, which has not yet been studied in the fracton literature.  (Earlier work\cite{foliated} has studied a field theory for a \emph{condensate} of ungauged fractons, though as we will discuss, such field theories cannot describe the mobility restrictions of uncondensed fractons.)  We therefore begin by formulating a field-theoretic description of ungauged fractons exhibiting global conservation of both charge and dipole moment.  We show that there is generically no nontrivial Gaussian action ($i.e.$ quadratic in the fracton fields) with these properties.  Rather, the natural field theory for the simplest fracton system is \emph{quartic} in the fracton fields.  Similarly, field theories for fracton systems conserving even higher moments will feature even higher powers of the fields in the action.  This type of non-Gaussian behavior has already been found by Slagle and Kim in the context of a gauge theory for the ``X-cube" fracton model.\cite{field}  This non-Gaussian nature of the field theory reflects the fact that fractons necessarily interact with each other even in the absence of a mediating gauge field, as we will review.  In this sense, there is no true ``non-interacting" theory of fractons.

With the ungauged theory in hand, we then proceed to apply the gauge principle, demanding that the theory be invariant under local symmetry transformations.  We show that, for a theory conserving charge and dipole moment, the gauge principle demands coupling to a rank-two symmetric tensor gauge field, consistent with previously studied fracton phases.\cite{sub}  Similarly, local conservation of higher charge moments will require the introduction of even higher rank tensor gauge fields.  In this way, the theory of symmetric tensor gauge fields can be derived directly from a gauge principle, in close analogy with more conventional gauge theories.  Finally, we consider extensions of this logic to other types of subdimensional particles besides fractons, which also yield tensor gauge theories upon application of the gauge principle.  This work opens a possible door to investigations of non-abelian fracton theories, via gauging non-abelian analogues of higher moment symmetries.

\emph{Ungauged Fracton Field Theory}.  We begin by constructing a field theory for ungauged fractons, focusing on the simplest type: $U(1)$ fractons exhibiting conservation of both charge and dipole moment.  We describe the fractons by a matter field $\Phi$ along with its corresponding charge density operator, $\rho = \Phi^\dagger\Phi$.  We now wish to write down an action for this theory which is consistent with global conservation of charge and dipole moment.  Since the charge density $\rho$ generates rotations of the phase of $\Phi$, the demand that the action respect charge conservation ($\int d^dx\,\rho = \textrm{constant}$) requires invariance of the action under the global transformation:
\begin{equation}
\Phi\rightarrow e^{i\alpha}\Phi
\end{equation}
for constant $\alpha$.  We also demand that the theory obeys conservation of dipole moment, $\int d^dx\,(\rho\vec{x}) = \textrm{constant}$, which requires an additional invariance under:
\begin{equation}
\Phi\rightarrow e^{i\vec{\lambda}\cdot \vec{x}}\Phi
\end{equation}
for constant vector $\vec{\lambda}$.  In other words, the phase of the field can now change by a linear function, instead of simply by a global constant, as in the case of ordinary conserved charges.  For convenience, we combine both types of transformations into the form:
\begin{equation}
\Phi\rightarrow e^{i\alpha(x)}\Phi
\end{equation}
where, in the present context, $\alpha(x)$ is restricted to be a linear function.

In order to construct an action with the desired symmetries, we look for operators $O$ which transform covariantly under the phase rotation, $O\rightarrow e^{in\alpha}O$ for integer $n$.  For ordinary globally conserved charges, where $\alpha(x)$ is a constant, the field $\Phi$ and all of its derivatives and powers transform covariantly.  In the present case, however, the linear behavior of $\alpha(x)$ restricts the set of covariant operators involving spatial derivatives.  While the field $\Phi$ itself is still covariant, it can readily be checked that any number of derivatives acting on a single power of $\Phi$ will not transform covariantly.  Rather, the lowest-order covariant derivative operator contains \emph{two} factors of $\Phi$, taking the form:
\begin{equation}
\Phi\partial_i\partial_j\Phi - \partial_i\Phi\partial_j\Phi
\end{equation}
Under a generic transformation $\Phi\rightarrow e^{i\alpha(x)}\Phi$, this operator transforms into:
\begin{align}
\begin{split}
e^{i2\alpha}\bigg(\Phi\partial_i\partial_j\Phi -& \partial_i\Phi\partial_j\Phi \\
 + i\partial_i\alpha\Phi\partial_j\Phi& + i\partial_j\alpha\Phi\partial_i\Phi - (\partial_i\alpha\partial_j\alpha - i \partial_i\partial_j\alpha)\Phi^2 \\
-i\partial_i\alpha\Phi\partial_j\Phi& - i\partial_j\alpha\Phi\partial_i\Phi + \partial_i\alpha\partial_j\alpha\Phi^2\bigg) \\
= e^{i2\alpha}\bigg(&\Phi\partial_i\partial_j\Phi - \partial_i\Phi\partial_j\Phi + (i\partial_i\partial_j\alpha)\Phi^2\bigg)
\end{split}
\label{trans}
\end{align}
For the phase rotations under consideration, where $\alpha(x)$ is a linear function, we have $\partial_i\partial_j\alpha = 0$, such that:
\begin{equation}
\Phi\partial_i\partial_j\Phi - \partial_i\Phi\partial_j\Phi \rightarrow e^{i2\alpha}(\Phi\partial_i\partial_j\Phi - \partial_i\Phi\partial_j\Phi)
\end{equation}
It is then straightforward to construct a Lagrangian respecting the charge and dipole conservation laws.  To lowest order, this Lagrangian takes the form:
\begin{align}
\begin{split}
\mathcal{L} = |\partial_t\Phi|^2 - m^2|\Phi|^2 - \lambda|\Phi\partial_i\partial_j\Phi - \partial_i\Phi\partial_j\Phi|^2 \\
-\lambda'\Phi^{*2}(\Phi\partial^2\Phi - \partial_i\Phi\partial^i\Phi)
\label{field}
\end{split}
\end{align}
where we have freely added a term with a single time derivative and a mass term, which have no interplay with the spatially dependent phase rotation.  The constants $\lambda$ and $\lambda'$ are arbitrary couplings.  Note that, while the $\lambda'$ term contains fewer derivatives than the $\lambda$ term, it only contains diagonal second derivatives ($e.g.$ $\partial_x^2$, but not $\partial_x\partial_y$), indicating that this term only describes longitudinal motion of dipoles.  As such, it is necessary to keep the $\lambda$ term in order to correctly describe transverse motion of dipoles.  Also note that we have here assumed rotational invariance, for simplicity.  Additional anisotropic terms may arise for certain lattice symmetries.

\begin{figure}[t!]
 \centering
 \includegraphics[scale=0.35]{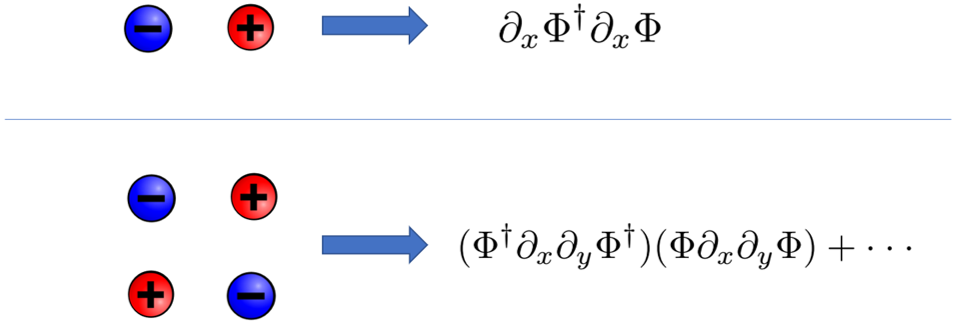}
 \caption{Theories with only charge conservation have local dipole creation operators, which lead to quadratic terms in the continuum field theory limit.  In contrast, a theory with dipole moment conservation has no such dipole creation operators.  In this case, the smallest creation operators are quadrupolar, leading to quartic terms in the field theory.}
 \label{fig:creation}
\end{figure}

Equation \ref{field} represents a field theory for ungauged fractons, respecting global conservation of both charge and dipole moment.  Importantly, this lowest-order nontrivial action is already non-Gaussian before coupling to any gauge field, containing fourth powers of $\Phi$ in the action.  This non-Gaussian behavior could have been anticipated based on microscopic considerations.  For a theory of ordinary conserved charges, local charge creation operators take the form of microscopic dipoles, $\Phi^\dagger(x+1)\Phi(x)$, which lead to a quadratic term in the action upon Taylor expansion.  Meanwhile, in a theory with conserved dipole moment, the local charge creation operators take the form of microscopic quadrupoles, as depicted in Figure \ref{fig:creation}, leading to quartic terms upon Taylor expansion.  Similarly, a theory with conserved quadrupole moment would only have microscopic octupole operators, leading to an action which is octic in the fracton field.  This logic will extend to conservation laws of any higher moment.

\begin{figure}[t!]
 \centering
 \includegraphics[scale=0.32]{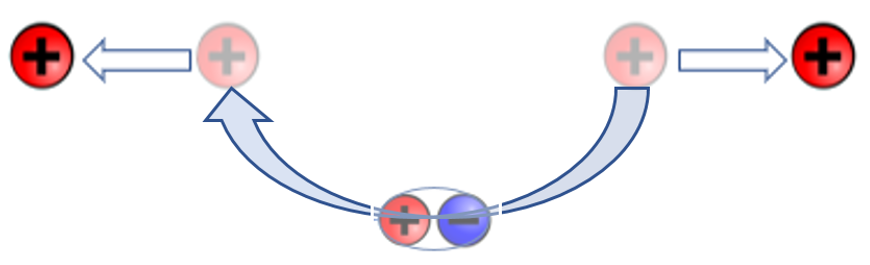}
 \caption{Fractons can interact via the exchange of virtual dipoles, allowing two fractons to ``push off" of each other.}
 \label{fig:interact}
\end{figure}

The non-Gaussian nature of the fracton action is also to be expected, since fractons have an intrinsic ability to interact with each other, even in the absence of a mediating gauge field.  While a fracton cannot move by itself, a fracton is capable of limited mobility by ``pushing off" other fractons in the system via the following process:  A fracton moves in one direction by emitting a dipole in the opposite direction, which then propagates to and is absorbed by a second fracton, as depicted in Figure \ref{fig:interact}.  Such processes, which lead to a ``gravitational" attraction between fractons\cite{mach}, are neatly captured by the two quartic terms in the fracton action.

We mention in passing that, to describe a symmetry broken system, such that $\Phi(x,t) = \Phi_0e^{i\theta(x,t)}$ for constant $\Phi_0$ and dynamical phase $\theta$, the Lagrangian of Equation \ref{field} will simplify to an ordinary Gaussian field theory on the phase $\theta$, which transforms as $\theta\rightarrow \theta + \alpha(x)$ under the symmetries.  The resulting action takes the form:
\begin{equation}
\mathcal{L} = \frac{1}{2}(\partial_t\theta)^2 - \frac{K}{2}(\partial_i\partial_j\theta)^2
\label{symbreak}
\end{equation}
for some parameter $K$.  For such symmetry broken phases, where the mobility restrictions have been relaxed, the non-Gaussian nature of the original action is unimportant.  However, for describing a symmetric system with immobile gapped fractons, one must retain the full quartic structure of the field theory.  Note that, if $(\partial_i\partial_j\theta)^2$ is replaced by $\cos(\partial_i\partial_j\theta)$ in the above Lagrangian, then accessing the uncondensed phase becomes possible.  However, such cosine terms inherently require a choice of underlying lattice and do not represent a true continuum field theory.  Rather, the quartic action of Equation \ref{field} must be used for a completely continuum description of the fracton phase.

\emph{Application of the Gauge Principle}.  In the previous section, we derived a fracton field theory which is invariant under $\Phi\rightarrow e^{i\alpha(x)}\Phi$ for linear functions $\alpha(x)$, thereby respecting global conservation of charge and dipole moment.  We now wish to apply the gauge principle and demand that the theory have a local symmetry, such that $\alpha(x)$ is an arbitrary function of $x$.  From the form of the transformation seen in Equation \ref{trans}, it is easy to see that we can construct a gauge-covariant object of the form:
\begin{equation}
\Phi\partial_i\partial_j\Phi - \partial_i\Phi\partial_j\Phi - iA_{ij}\Phi^2
\label{object}
\end{equation}
if we have the transformation rules:
\begin{align}
&\Phi\rightarrow e^{i\alpha}\Phi \\
A_{ij}&\rightarrow A_{ij} + \partial_i\partial_j\alpha
\end{align}
For notational simplicity, we define a gauge-covariant second derivative operator $D_{ij}$ acting on $\Phi^2$ as follows:
\begin{equation}
D_{ij}\Phi^2 = \Phi\partial_i\partial_j\Phi - \partial_i\Phi\partial_j\Phi - iA_{ij}\Phi^2
\end{equation}
More properly, we should define $D_{ij}$ as a \emph{bilinear} operator acting on two functions $\Phi$ and $\Psi$ as $D_{ij}[\Phi,\Psi] = \Phi\partial_i\partial_j\Psi - \partial_i\Phi\partial_j\Psi - iA_{ij}\Phi\Psi$, where both fields must have the same charge under the gauge transformation in order for the operator to be covariant.  (Note that, in the presence of multiple charged fields, only the \emph{total} charge and dipole moment of the system will be conserved, not for each field separately.)  For present purposes, however, we will only need the ``diagonal" element, $D_{ij}\Phi^2\equiv D_{ij}[\Phi,\Phi]$.  We can also allow $\alpha$ to have arbitrary time dependence if we introduce the gauge-covariant time derivative:
\begin{equation}
D_t\Phi = (\partial_t - i \phi)\Phi
\end{equation}
where the field $\phi$ transforms as:
\begin{equation}
\phi\rightarrow \phi + \partial_t\alpha
\end{equation}
behaving analogously to a ``timelike" component of the gauge field.  We can also construct gauge-invariant quantities involving only the gauge fields, analogous to the electric and magnetic fields of ordinary $U(1)$ gauge theory.  For example, in three spatial dimensions these take the form $E_{ij} = \partial_tA_{ij} - \partial_i\partial_j\phi$ and $B_{ij} = \epsilon_{ik\ell}\partial^kA^\ell_{\,\,\,j}$.  Making use of these quantities, the Lagrangian for the gauged theory, to lowest order in derivatives, must take the form:
\begin{align}
\begin{split}
\mathcal{L} = |D_t\Phi|^2 - m^2|\Phi|^2 - \lambda|D_{ij}\Phi^2|^2 - \lambda'(\Phi^{*2}D^i_i\Phi^2+\textrm{c.c.})\\
-\lambda' + E^{ij}E_{ij} - B^{ij}B_{ij}
\end{split}
\end{align}
Note the need for the complex conjugate (c.c.) in the $\lambda'$ term, which is not automatically real-valued.  The gauge field sector of this Lagrangian is exactly of the form studied in previous works on fracton tensor gauge theories, specifically that of the scalar charge theory studied in Reference \onlinecite{sub}.  However, we now have an explicit coupling to matter fields which accounts for the higher moment conservation laws of fractons.  Through similar application of the gauge principle to theories with even higher conserved moments, we could also derive gauge theories featuring tensor gauge fields of rank higher than two, which we do not carry through here.

\emph{Extensions to Other Subdimensional Particles}.  We now wish to extend these ideas to other types of subdimensional particles, which have restricted mobility only in certain directions.  As a concrete example, we will focus on one-dimensional particles, restricted to motion along a line.  A similar analysis will hold for two-dimensional particles.  We focus on the simplest type of one-dimensional particles, which carry a vector-valued charge $\vec{\rho}$.  We take these vector charges to exhibit global conservation of both charge ($\int d^dx\,\vec{\rho}=\textrm{constant}$) and angular charge moment ($\int d^dx\,(\vec{\rho}\times\vec{x}) = \textrm{constant}$).  We describe these vector particles via a field $\Phi_i$ corresponding to each component of the charge vector, such that $\rho_i = \Phi^\dagger_i\Phi_i$, where no summation over $i$ is implied.  In this section alone, all summations will be indicated explicitly.

Conservation of vector charge implies that the action for the theory should be invariant under independent phase rotations on each component of the field:
\begin{equation}
\Phi_i\rightarrow e^{i\alpha_i}\Phi_i
\end{equation}
for constant vector $\alpha_i$, where this equation is to be interpreted component-wise in some particular basis.  Meanwhile, the angular moment conservation law implies invariance under a second set of transformations:
\begin{equation}
\Phi_i\rightarrow e^{i\sum_{jk}\epsilon_{ijk}\lambda_jx_k}\Phi_i
\end{equation}
for constant vector $\lambda^j$.  For convenience, we combine these transformations as:
\begin{equation}
\Phi_i\rightarrow e^{i\alpha_i(x)}\Phi_i
\end{equation}
where for present purposes we have $\alpha_i(x) = \alpha_{0,i} + \sum_{jk}\epsilon_{ijk}\lambda_jx_k$ for constants $\alpha_{0,i}$ and $\lambda$.  Note that the field $\Phi_i$ does not transform nicely under rotations ($i.e.$ it is not a valid vector).  Nevertheless, at the end of the day, we will obtain a true tensor gauge field upon applying the gauge principle.  The loss of manifest rotational invariance introduced by $\Phi_i$ may simply be a mathematical artifact.  It remains an open problem whether this theory can be rewritten in a manifestly rotationally invariant way.

We now wish to construct an invariant action by identifying the covariant operators in this theory.  As usual, operators without derivatives are automatically covariant.  Meanwhile, a derivative on $\Phi_i$ transforms as:
\begin{equation}
\partial_i\Phi_j\rightarrow e^{i\alpha_j}(\partial_i + i\partial_i\alpha_j)\Phi_j = e^{i\alpha_j}(\partial_i + i\sum_k\epsilon_{ikj}\lambda_k)\Phi_j
\end{equation}
We see that, at the level of single-field derivative operators, we have covariance only when $i=j$, giving us the covariant longitudinal derivative operators:
\begin{equation}
\partial_i\Phi_i\rightarrow e^{i\alpha_i}\partial_i\Phi_i
\end{equation}
for each component $i$.  To include transverse derivatives in the action, however, we must once again proceed to two-field operators.  We can easily construct a covariant operator as follows:
\begin{align}
\begin{split}
&\Phi_i\partial_i\Phi_j + \Phi_j\partial_j\Phi_i\rightarrow \\
e^{i(\alpha_i+\alpha_j)}\bigg(\Phi_i\partial_i\Phi_j& + \Phi_j\partial_j\Phi_i + i(\partial_i\alpha_j + \partial_j\alpha_i)\Phi_i\Phi_j\bigg)
\end{split}
\end{align}
Taking $\alpha_i(x) = \alpha_{0,i} + \sum_{jk}\epsilon_{ijk}\lambda_jx_k$, the last term above vanishes, leaving us with:
\begin{equation}
\Phi_i\partial_i\Phi_j + \Phi_j\partial_j\Phi_i\rightarrow e^{i(\alpha_i+\alpha_j)}\bigg(\Phi_i\partial_i\Phi_j + \Phi_j\partial_j\Phi_i\bigg)
\end{equation}
We can now write down a Lagrangian respecting vector charge and angular moment conservation as a function of these covariant operators:
\begin{align}
\mathcal{L}[\Phi_i,\partial_i\Phi_i,\Phi_i\partial_i\Phi_j + \Phi_j\partial_j\Phi_i]
\label{vecact}
\end{align}
where the lowest order term involving transverse spatial derivatives is quartic in the fields.

Starting from this theory with global conservation laws, we can now apply the gauge principle, giving $\alpha_i(x,t)$ arbitrary spacetime dependence, to promote the conservation laws to local ones.  In this case, the gauge-covariant operators become:
\begin{equation}
\Phi_i\partial_i\Phi_j + \Phi_j\partial_j\Phi_i - iA_{ij}\Phi_i\Phi_j
\end{equation}
\begin{equation}
(\partial_i - iA_{ii})\Phi_i
\end{equation}
\begin{equation}
D_t\Phi_i = (\partial_t - i\phi_i)\Phi_i
\end{equation}
where we have introduced a tensor gauge field which transforms as:
\begin{equation}
A_{ij}\rightarrow A_{ij}+\partial_i\alpha_j + \partial_j\alpha_i
\end{equation}
\begin{equation}
\phi_i\rightarrow\phi_i + \partial_t\alpha_i
\end{equation}
The corresponding action can be written down directly from Equation \ref{vecact} by replacing all derivative operators with their gauge-covariant versions, plus adding the appropriate field invariants.  The resulting gauge theory has exactly the structure of the vector charge theory discussed in Reference \onlinecite{sub}.  In this way, tensor gauge theories of subdimensional particles can be derived from a gauge principle, just like those for fractons.

\emph{Conclusions}.  In this work, we applied the ``gauge principle" ($i.e.$ promotion of a global symmetry to a local symmetry) to a system of fractons.  We have shown how to formulate field theories of ungauged fractons, obeying global conservation of higher charge moments.  These field theories are generically non-Gaussian, reflecting the intrinsic ability of fractons to interact without the need for a mediating gauge field.  We then promoted the global higher moment conservation laws to local ones, which we have shown requires coupling the theory to a symmetric tensor gauge field.  In this way, the theory of symmetric tensor gauge fields arises from a gauge principle in much the same way as an ordinary vector gauge field.  We also discussed extensions of this logic to other types of subdimensional particles besides fractons, which obey similar gauge principles.

This work has the potential to open several new directions in fracton physics.  For example, by considering a fracton field transforming under a nonabelian Lie group, is it possible to construct a nonabelian tensor gauge theory?  And what would the properties be of such a system?  Also, with an explicit field theory for fractons, is it now possible to develop more powerful technical tools to analyze fracton theories, such as an analogue of Feynman diagrams?  By discretizing the derivatives, can we write fracton theories on arbitrary lattices?  How do these field theories need to be modified to apply to ``type-II" $U(1)$ fractons, featuring no mobile bound states?\cite{uhaah}  Does this field theory shed any light on the theory of elasticity and its associated phase transitions?\cite{elasticity,pai,gromov,supersolid}  There are many exciting questions remaining to be answered.

\emph{Acknowledgments}.  I acknowledge useful conversations with Wojciech de Roeck, Mike Hermele, Shriya Pai, Leo Radzihovsky, Rahul Nandkishore, Han Ma, Abhinav Prem, and Kevin Slagle.  I also thank Juven Wang for pointing out some typos in a previous version of this manuscript.  This work is supported partially by NSF Grant 1734006, by a Simons Investigator Award to Leo Radzihovsky, and by the Foundational Questions Institute (fqxi.org; grant no. FQXi-RFP-1617) through their fund at the Silicon Valley Community Foundation.

\end{document}